\documentclass[aps,amssymb,showpacs]{revtex4}
\usepackage{graphicx}
\usepackage{longtable}
\usepackage{epsfig}
\usepackage{dcolumn}
\usepackage{bm}

\begin{document}

\title{Order parameters in the Verwey phase transition}

\author{      Przemys\l{}aw Piekarz}
\affiliation{ Institute of Nuclear Physics, Polish Academy of Sciences, 
              Radzikowskiego 152, PL-31342 Krak\'{o}w, Poland }

\author{      Krzysztof Parlinski }
\affiliation{ Institute of Nuclear Physics, Polish Academy of Sciences, 
              Radzikowskiego 152, PL-31342 Krak\'{o}w, Poland }

\author{      Andrzej M. Ole\'{s} }
\affiliation{ Institute of Nuclear Physics, Polish Academy of Sciences, 
              Radzikowskiego 152, PL-31342 Krak\'{o}w, Poland }

\begin{abstract}
The Verwey phase transition in magnetite is analyzed on the basis
of the Landau theory. The free energy functional is expanded in a
series of components belonging to the primary and secondary
order parameters. A low-temperature phase with the monoclinic P2/c
symmetry is a result of condensation of two order
parameters $X_3$ and $\Delta _5$. The temperature dependence of
the shear elastic constant $C_{44}$ is derived and the mechanism
of its softening is discussed.
\end{abstract}


\maketitle

\section{Introduction}

The mechanism of the phase transition in magnetite (Fe$_3$O$_4$)
at $T_V=122$ K, discovered by Verwey \cite{Verwey}, has remained a
big puzzle in the condensed matter physics for almost 70 years.
Developments in experimental and theoretical methods during last
years enabled to reveal subtle changes in the crystal and electronic
structure below $T_V$ \cite{Attfield,LDAU}. A simple charge
ordering picture in which metal-insulator transition is induced by
electrostatic interactions was replaced by a highly complex
scenario in which lattice, charge, spin and orbital degrees of
freedom are involved. Recent theoretical studies revealed the
important role of the local electron interactions and orbital
correlations in the $t_{2g}$ states on iron ions \cite{LDAU}.

\begin{figure}[b!]
\begin{center}
\includegraphics[width=7.7cm]{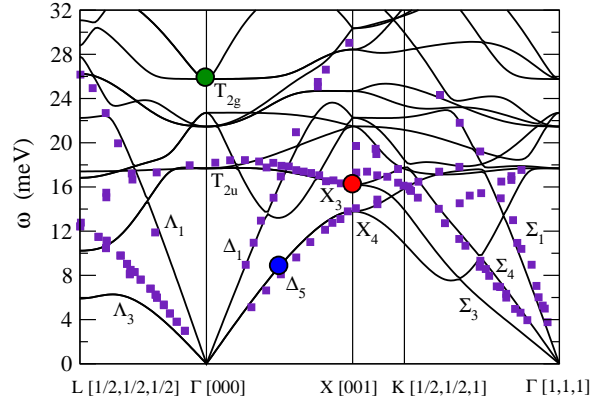}
\end{center}
\caption{The phonon dispersion curves of Fe$_3$O$_4$ (solid lines) compared with
the experimental points taken from Ref. \cite{neutron}. The order parameters $X_3$, $\Delta_5$,
and $T_{2g}$ are marked by circles.}
\end{figure}

The electronic interactions are complemented by the lattice
deformation, which breaks the cubic symmetry and induces a
low-temperature (LT) monoclinic phase driven by the
electron-phonon interactions \cite{Iizumi}. In the previous work
\cite{PRL,PRB}, we have obtained the phonon spectrum of magnetite
(Fig. 1) using the {\it ab initio} computational technique
\cite{direct}. We have identified two primary order parameters
(OPs) at ${\bf k}_{X}=\frac{2\pi}{a}(0,0,1)$ and ${\bf
k}_{\Delta}=\frac{2\pi}{a}(0,0,\frac{1}{2})$ with the $X_3$ and
$\Delta_5$ symmetry, respectively, which both play important role
in the VT: ($i$) the $\Delta_5$ mode is responsible for the
doubling of the unit cell along the $c$ direction in the
monoclinic phase, while ($ii$) the $X_3$ phonon induces the
metal-insulator transition by its coupling to the electronic
states near the Fermi energy \cite{PRL}. Due to the
electron-phonon interaction the above OPs are combinations of the
electron (charge-orbital) and lattice components. This explains
why the phonon soft mode has not been observed. Instead,
low-energy critical fluctuations of OPs were found by the diffuse
neutron scattering \cite{Diffuse}. The condensation of the OPs
below $T_V$ explains the crystal symmetry change as well as the
charge-orbital ordering.

The group theory predicts also secondary OPs, which do not effect
the symmetry below $T_V$ but modify the properties of magnetite
close to a transition point. At the $\Gamma$ point, the $T_{2g}$
mode can be classified as the secondary OP, and its coupling to
the shear strain explains the softening of the $C_{44}$ elastic
constant \cite{c44,Seikh}. The lowest $T_{2g}$ optic mode, marked in Fig.
1, could contribute quantitatively to the free energy, but it does
not play any significant role for the VT.

In this work, we introduce and analyze the Landau free energy for
the VT, and discuss a solution corresponding to the LT monoclinic
phase. We derive also the temperature dependence of $C_{44}$.

\section{Free energy}

The Landau free energy can be expanded into a series of the
components of the OPs. The invariant terms describing couplings
between the OPs were derived using the group theory methods
\cite{Iso}. The only nonzero components of the primary OPs $X_3$
and $\Delta_5$ are denoted by $g$ and $q$, respectively. We
include also the secondary OP with the $T_{2g}$ symmetry ($\eta$)
and shear-strain ($\epsilon$). The free energy can be written in
the form \cite{PRB}
\begin{equation}
{\cal F}={\cal F}_0+\frac{\alpha_1}{2}g^2+\frac{\beta_1}{4}g^4+\frac{\gamma_1}{6}g^6
+\frac{\alpha_2}{2}q^2+\frac{\beta_2}{4}q^4+\frac{\delta_1}{2}g^2q^2
+\frac{\alpha_3}{2}\eta^2+\frac{\alpha_4}{2}\epsilon^2+\frac{\delta_2}{2}\eta g^2
+\frac{\delta_3}{2}\epsilon g^2+\delta_4\eta\epsilon,
\label{free}
\end{equation}
were ${\cal F}_0$ is the part of the potential, which does not
change through the transition. We assume that $\beta_1>0$,
$\beta_2>0$ and $\gamma_1>0$ to ensure the stability of the
potential at high temperatures. For the second-order terms we
assume standard temperature behavior $\alpha_i=a_i(T-T_{ci})$ near
the critical temperature $T_{ci}$ for $i=1,2,3$, which would
correspond to a continuous phase transition.
The coefficient $\alpha_4$ is the shear elastic constant
at high temperatures ($C^0_{44}$). The coupling between the primary
OPs is biquadratic, between the secondary and primary OPs has the
linear-quadratic form, and the coupling between the components of
the secondary OP is of the bilinear type. Taking first derivatives
of ${\cal F}$ over all OPs we get
\begin{eqnarray}
\frac{\partial {\cal F}}{\partial g}&=&g(\alpha_1+\beta_1g^2+\gamma_1g^4+\delta_1q^2+\delta_2\eta+\delta_3\epsilon)=0, \\
\frac{\partial {\cal F}}{\partial q}&=&q(\alpha_2+\beta_2q^2+\delta_1g^2)=0, \\
\frac{\partial {\cal F}}{\partial \eta}&=&\alpha_3\eta+\delta_4\epsilon+\frac{\delta_2}{2}g^2=0, \\
\frac{\partial {\cal F}}{\partial \epsilon}&=&\alpha_4\epsilon+\delta_4\eta+\frac{\delta_3}{2}g^2=0.
\end{eqnarray}
The solution $g=q=\eta=\epsilon=0$ corresponds to the
high-temperature cubic symmetry ($Fd\bar{3}m$). 
From Eq. (3) we obtain the dependence between $g$ and $q$
\begin{equation}
q^2=-\frac{\delta_1g^2+\alpha_2}{\beta_2},
\label{qg}
\end{equation}
which has three possible solutions: ($i$) $g=0$ and
$q^2=-\frac{\alpha_2}{\beta_2}$ if $\alpha_2<0$ ($Pbcm$), ($ii$)
$q=0$ and $g^2=-\frac{\alpha_2}{\delta_1}$ if $\alpha_2>0$ and
$\delta_1>0$ or $\alpha_2<0$ and $\delta_1>0$ ($Pmna$), ($iii$)
$g\neq 0$ and $q\neq 0$ ($P2/c$). In the brackets we put the space
group symbols, which characterize the low-symmetry phases. The solution
($iii$) which corresponds to the experimentally observed LT
monoclinic phase requires simultaneous condensation of both
primary OPs. The necessary condition for this is a negative value
of $\delta_1$. Indeed, it has been established by the {\it ab
initio} studies that the total energy is lowered when the crystal
is distorted by both $X_3$ and $\Delta _5$ modes \cite{PRB}. For
$\delta_1<0$, Eq. (\ref{qg}) has a non-zero solution provided that
$|\delta_1|g^2>\alpha_2$. It implies that for $\alpha_2>0$
($T>T_{c2}$), the phase transition occurs when the OP $g$ exceeds
a critical value $\frac{\alpha_2}{|\delta_1|}$, so it has a
discontinuous (first-order) character.

From Eqs. (4) and (5) we get
\begin{equation}
\eta=\frac{\delta_3\delta_4-\delta_2\alpha_4}{2\alpha_3\alpha_4
-2\delta^2_4}g^2\equiv\lambda_1 g^2, \hskip 1.5cm
\epsilon=\frac{\delta_2\delta_4-\delta_3\alpha_3}{2\alpha_3\alpha_4
-2\delta^2_4}g^2\equiv\lambda_2g^2,
\end{equation}
which shows that $\eta\neq0$ and $\epsilon\neq0$ only if $g\ne0$.
Eliminating $q$, $\eta$ and $\epsilon$ using  Eqs. (6) and (7),
the potential ${\cal F}$ can be written as a function of $g$
\begin{equation}
{\cal F}={\cal F}_0'
+\frac{\alpha}{2}g^2+\frac{\beta}{4}g^4+\frac{\gamma_1}{6}g^6,
\end{equation}
where the renormalized coefficients are
\begin{eqnarray}
{\cal F}_0'&=&{\cal F}_0-\frac{\alpha^2_2}{4\beta_2}, \\
\alpha&=&\alpha_1-\frac{\alpha_2\delta_1}{\beta_2},   \\
\beta&=&\beta_1-\frac{\delta^2_1}{\beta_2}+2\alpha_3\lambda^2_1+2\alpha_4\lambda^2_2+
2\delta_2\lambda_1+2\delta_3\lambda_2+4\delta_4\lambda_1\lambda_2.
\end{eqnarray}
The zero-order and second-order terms depend on the parameters
belonging to the primary OPs. The secondary OPs modify only the
forth-order term. In this notation, the solution of Eqs. (2)-(5),
which minimizes the potential ${\cal F}$ reads
\begin{equation}
g_o^2=\frac{-\beta+\sqrt{\beta^2-4\gamma\alpha}}{2\gamma}, \hskip
.7cm q^2_o=-\frac{\delta_1g^2_o+\alpha_2}{\beta_2}, \hskip .7cm
\eta_o=\lambda_1g^2_o, \hskip .7cm \epsilon_o=\lambda_2g^2_o.
\end{equation}

To study the softening of $C_{44}$, we have eliminated $g$, $q$,
and $\eta$ using Eqs. (2), (3), and (4), and expressed the free
energy as a function of $\epsilon$ only. In these calculations we
have omitted the sixth-order term, which usually has a small
contribution near the transition point. The elastic constant
$C_{44}$ is obtained using the standard definition
\begin{equation}
C_{44}(T)=\frac{\partial^2 {\cal
F}}{\partial\epsilon^2}=C^0_{44}-\frac{\delta^2}{\alpha'_3}
-\frac{\delta^2_3}{\beta'_1}, \label{c44}
\end{equation}
where
\begin{equation}
\delta=\delta_4-\frac{\delta_2\delta_3}{2\beta'_1},\hskip .7cm
\alpha'_3=\alpha_3-\frac{\delta^2_2}{2\beta'_1}=a_3(T-T'_{c3}),
\hskip .7cm \beta'_1=\beta_1-\frac{\delta^2_1}{\beta_2},
\end{equation}
with $T'_{c3}=T_{c3}+\delta^2_2/2\beta'_1$. The second and third
term in Eq. (\ref{c44}) are negative at high temperatures, so both
contribute to the softening of $C_{44}$. It means that all
couplings included in Eq. (\ref{free}) are involved in this
behavior. The main temperature dependence is caused by the second
term, but also the last term in Eq. (\ref{c44}) may depend on
temperature. In Ref. \cite{c44}, the softening of $C_{44}$ was
explained taking into account only the bilinear coupling
($\delta_4$). Indeed, if we assume that $\delta_2=\delta_3=0$, Eq.
(\ref{c44}) reduces to that found in Ref. \cite{c44}:
$C_{44}=C^0_{44}-\delta^2_4/\alpha_3$. Omitting the last term, Eq.
(\ref{c44}) can be written in the form
\begin{equation}
C_{44}=C^0_{44}\;\frac{T-T_0}{T-T'_{c3}}, \label{CT}
\end{equation}
where $T_0=T'_{c3}+\delta^2/C^0_{44}a_3$. Since, Eq. (\ref{CT})
has the same form as that one discussed in Ref. \cite{c44}, the
fitting procedure will give the following values of parameters:
$T_0=66$ K and $T'_{c3}=56$ K. Note that the meaning of both
temperatures $T_0$ and $T'_{c3}$ is here different than in Ref.
\cite{c44} as they include all interactions. The softening is not
complete (only about 10 \%) since the first-order phase transition
and the change of structure occurs at much higher temperature than
$T_0$.

\section{Discussion}

The present work shows that the VT can be analyzed using the
Landau theory of phase transitions. It provides a basis to study
the interplay between the OPs and the mechanism of phase
transition with two or more OPs. Moreover, majority of
experimental facts can be understood in one coherent picture. We
emphasize that the coupling between the modes with the $X_3$ and
$\Delta_5$ symmetry plays the crucial role in the VT in
magnetite and explains both the occurrence of the LT monoclinic
phase and the metal-insulator transition. This results in rather
complex structure of the charge and orbital ordering, with two
different charge modulations characterized by  ${\bf k}_X$ and
${\bf k}_{\Delta}$ wave vectors found in diffraction studies
\cite{Attfield}. In this work, we have discussed the conditions,
which have to be fulfilled for both OPs to develop simultaneously.
It depends primarily on the coupling coefficient $\delta_1$, which
should be negative below $T_V$ and large enough to stabilize both
OPs.

Our present analysis extends also the previous studies of the
temperature dependence of $C_{44}$. So far, the softening of
$C_{44}$ was studied in the models restricted to the $\Gamma$
point \cite{c44,Seikh}. Since the phase transition is driven
mainly by the OPs at ${\bf k}\neq 0$, the OPs at the zone center
develop solely due to their coupling to other modes. Therefore,
the bilinear coupling at the $\Gamma$ point is rather a side
effect of the phase transition, not the main origin of it. The
present study shows that also a direct coupling between the $X_3$
primary OP and the shear strain ($\delta_3$) influences the
$C_{44}$ elastic constant, providing an alternative mechanism of
its softening.

\

This work was supported in part by Marie Curie Research Training
Network under Contract No. MRTN-CT-2006-035957 (c2c). 
A. M. Ole\'s acknowledges partial support by Foundation of Polish Science and
by the Polish Ministry of Science and Education
Project N202 068 32/1481.


\section{References}
\begin{thebibliography}{99}

\bibitem{Verwey} E J W Verwey 1939 Nature (London) {\bf 144} 327
\bibitem{Attfield} J P Wright, J P Attfield and P G Radaelli 2001
                   Phys. Rev. Lett. {\bf 87} 266401; \\
                 J P Wright, J P Attfield and P G Radaelli 2002
                   Phys. Rev. B {\bf 66} 214422
\bibitem{LDAU} I Leonov, A N Yaresko, V N Antonov, M A Korotin,
                 and V I Anisimov 2004 Phys. Rev. Lett. {\bf 93} 146404
\bibitem{Iizumi} M Iizumi, T F Koetzle, G Shirane, S Chikazumi, M Matsui
                 and S Todo 1982 Acta Cryst. B {\bf 38} 2121
\bibitem{PRL} P Piekarz, K Parlinski and A M Ole\'{s} 2006
                 Phys. Rev. Lett. {\bf 97} 156402
\bibitem{PRB} P Piekarz, K Parlinski and A M Ole\'{s} 2007
                 Phys. Rev. B {\bf 76} 165124 
\bibitem{direct} K Parlinski, Z Q Li and Y Kawazoe 1997
                 Phys. Rev. Lett. {\bf 78} 4063;\\
                 K. Parlinski 2005 {\sc phonon} Software
\bibitem{neutron} E J Samuelsen and O Steinsvoll 1974
                 Phys. Status Solidi B {\bf 61} 615
\bibitem{Diffuse} Y Fuji, G Shirane and Y Yamada 1975
                 Phys. Rev. B {\bf 11} 2036
\bibitem{c44} H Shwenk, S Bareiter, C Hinkel, B L\"{u}thi, Z Kakol,
              A Koz\l{}owski and J M Honig 2000
              Eur. Phys. J. B {\bf 13} 491
\bibitem{Seikh} M M Seikh, C Narayana, P A Metcalf, J M Honig
                and A K Sood 2005 Phys. Rev. B {\bf 71} 174106
\bibitem{Iso} H T Stokes and D M Hasch 2002 {\sc isotropy}
              software; stokes.byu.edu/isotropy.html

\end{thebibliography}
\end{document}